\documentclass{article}
\usepackage{spconf,amsmath,graphicx,hyperref,amssymb}
\usepackage{color}
\usepackage{float}

\def\bs{\mbox{\boldmath $s$}}

\def\b1{\mbox{\boldmath $1$}}
\def\b0{\mbox{\boldmath $0$}}

\def\my{\mbox{$\mathbf{y}$}}

\def\mB{\mbox{$\mathbf{B}$}}

\def\mD{\mbox{$\mathbf{D}$}}

\def\mI{\mbox{$\mathbf{I}$}}

\def\mL{\mbox{$\mathbf{L}$}}

\def\mU{\mbox{$\mathbf{U}$}}

\floatstyle{ruled}

\title{Topological Signal Processing for 3D Point Cloud Data }
\name{Tiziana Cattai$^1$, Stefania Sardellitti$^2$, Stefania Colonnese$^1$, Sergio Barbarossa$^1$ \thanks{This work was supported by the European Union - Next Generation EU under the Italian National Recovery and Resilience Plan (NRRP), Mission 4, Component 2, Investment 1.3, CUP  B53C22004050001, partnership on “Telecommunications of the Future” (PE00000001 - program “RESTART”).}}
\address{$^1$ Dept. of Information Engineering, Electronics and Telecommunications, \\Sapienza University of Rome, Italy,\\
$^2$Dept. of Engineering and Sciences, Universitas Mercatorum, Rome, Italy }
\begin{document}
%\ninept
%
\maketitle
\begin{abstract}
Our goal in this paper is to apply the topological signal processing (TSP) framework to the analysis of 3D Point Clouds (PCs) represented on  simplicial complexes. Building on Discrete Exterior Calculus (DEC) theory for vector fields, we introduce higher-order Laplacian operators that enable the processing of signals over triangular meshes. Unlike traditional approaches, the proposed approach allows us to characterize both color attributes, modeled as 3D vectors on nodes, and geometry, modeled as 3D vectors on the barycenter of each triangle.
Then, we show as TSP tools may efficiently be used to sample, recover and filter PCs attributes treating them as edge signals. 
Numerical results on synthetic PCs demonstrate accurate color reconstruction with robustness to sparse data and geometry refinement in the case of noisy PC coordinates. The proposed approach provides a topology-based representation to characterize the geometry and attributes of PCs.
\end{abstract}

\section{Introduction}
\label{sec:intro}
Point Clouds (PCs) are a set of points on a 3D surface and they are efficiently used to represent objects and environments in eXtended Reality (XR) applications. % to model objects and environments.
Each point of the PC is associated with its 3D coordinates, keeping its shape, together with their attribute, such as color or luminance \cite{graziosi2020overview}, \cite{cao20193d}.
A PC typically consists of a very large number of points so that processing operations are often costly and inefficient. 
To mitigate these issues, careful sampling strategies need to be designed to select an optimal subset, elaborate a compressed model, and subsequently rebuild the original structure \cite{sridhara2024graph}. In this context, the PC geometry plays a fundamental role in the characterization of objects and backgrounds as  in, e.g., XR applications. Inaccurate PC coordinates generate uncertain 3D models, and consequently, geometry refinement is a necessary task to improve PC quality and XR usability \cite{mattei2017point, irfan2021joint}.
Graph Signal Processing (GSP) tools \cite{shuman2013emerging,anis2014towards,7208894}, specifically developed to process signals defined over the nodes of a graph, have been applied to model and analyze  PC data \cite{hu2021graph}.  Graph-based representations fit well with PC, where each point has a set of features associated \cite{zhao2022graphreg}. Recently, the Topological Signal Processing (TSP)  framework \cite{barbarossa2020topological} has been developed for processing signals defined on higher order topological spaces than graphs, such as simplicial and cell complexes \cite{sardellitti2024topological,schaub2021signal}, to capture higher-order interactions, beyond pairwise relationships. In \cite{sardellitti2024topological}, TSP has also been applied to the analysis of planar vector fields, such as, for example, filtering RNA velocity vector fields.

\textbf{Contributions.} In this paper, we propose a TSP framework for PCs, where simplicial complexes are used to model higher-order structures. Based on DEC and Whitney interpolation functions, our framework enables a consistent mapping from vector signals defined on the vertices (color attributes) to scalar signals defined on the edges of a graph. Then, we associate a second order simplicial complex to the graph to exploit triple-wise interactions. The same framework is then extended to represent the PC geometry as vector fields defined on $2$-simplices (triangles). An example is given by the vectors associated to the normal of each triangle. This enables the analysis and refinement of noisy surface models. The main advantage of this approach is its ability to handle both missing attributes, such as color information, which may be absent due to compression or acquisition gaps, and uncertainties in the PC geometry, where noisy coordinates affect the surface normals. %We show the potential of the proposed framework on two problems: color recovery from sparse samples and geometry refinement under noisy conditions, showing improved reconstruction accuracy and robustness compared with graph-based methods.
The contributions of this paper can be summarized as follows: 1) we introduce an algorithmic generalization of the TSP framework to 3D PC data processing, modeling the PC as a simplicial complex able  to capture higher-order topological  relationships among data; 
     2) we define higher-order Laplacian operators, through DEC and Whitney interpolation functions, that enable the mapping from vector signals on the vertices to scalar signals on the edges, as well as the representation of vector fields on simplicial complexes;
3) we apply the proposed framework to two key problems in PC processing: (i) color recovery from sparse samples and (ii) geometry refinement under noisy point coordinates. We validate our approach on synthetic and real PCs.

 \section{From Point Cloud to simplicial complex} 
 
\begin{figure}
    \centering    \includegraphics[width=0.55\linewidth]{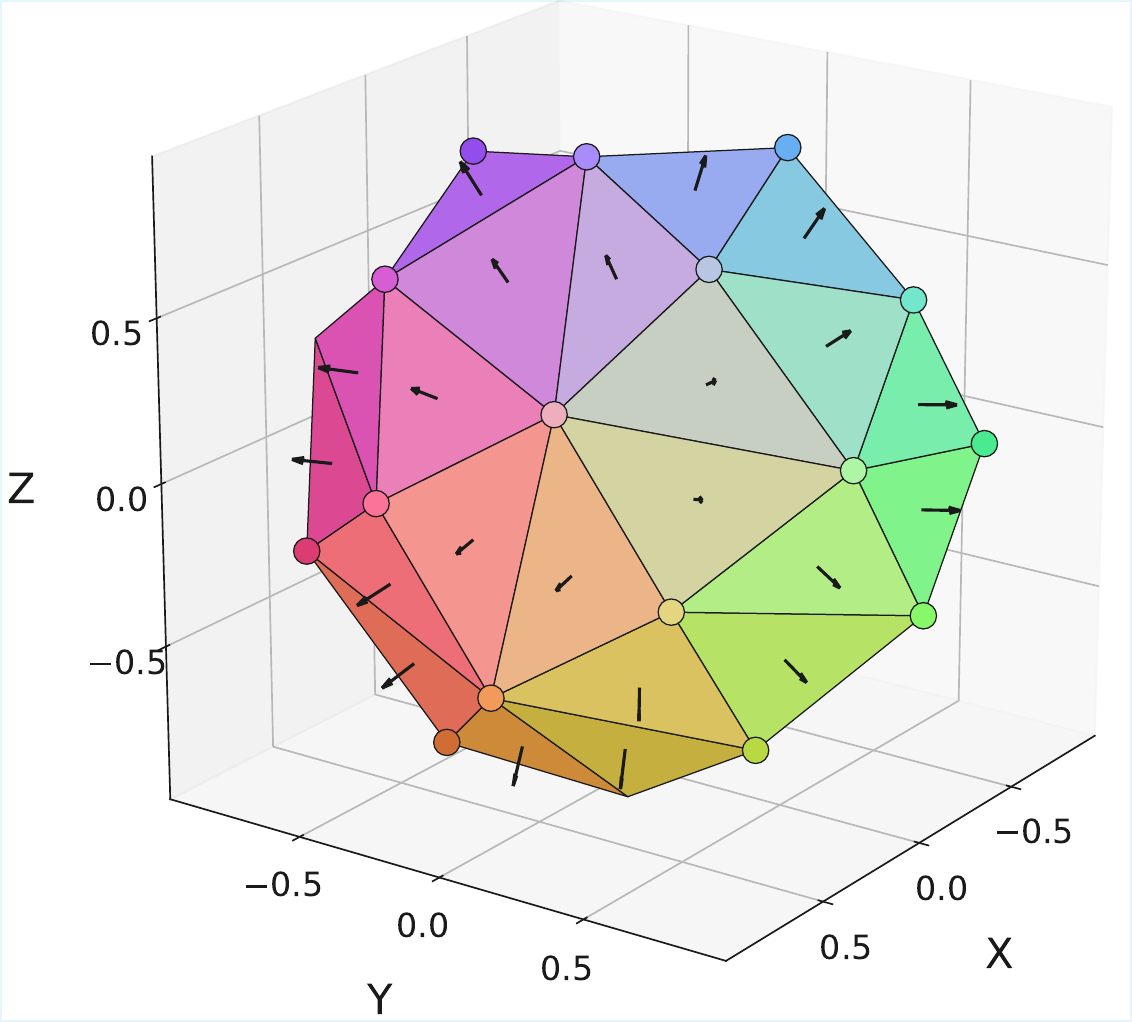}
    \caption{Example of geometric simplicial complex.}
    \label{fig:geom}
\end{figure}

By definition, a 3{D} point cloud $\mathcal{P}$ is a set of $N$ points $\mathcal{V}=\{v_i, \, i=1,\ldots,N\}$, embedded
in a 3{D} space and carrying both geometry and feature information \cite{cao20193d}.
The positions of the points are determined by the vectors $\mathbf{p}_i=[x_i,y_i,z_i]^T$, $i=1,\ldots,N$. 
Feature vectors are typically attached to each point to encode various attributes of the observed data. A typical example is the color vector
$(R,G, B)$.
%and normal vectors $(n_x ,n_y,n_z )$.
 %three-dimensional points $$\mathcal{V}=\left\{\textbf{v}_i\in  R^3,\; i=0,\cdots N-1\right\},$$ with  $ \mathbf{v_i}= [x_i, y_i,z_i]^T$. 
Various methods have been developed for generating triangular meshes out of the point cloud \cite{linsen2001point}. We build a triangular surface mesh $\mathcal{M}$, applying a well-centered Delaunay triangulation \cite{lee1980two}, \cite{chen2004optimal}.
Each triangle may have associated a feature vector, like, e.g., its normal vector $(n_x ,n_y,n_z )$. We denote with  $\mathcal{N}$, $\mathcal{E}$ and $\mathcal{T}$ the sets of nodes, edges and   triangles, respectively,  induced by the triangular mesh. The cardinalities  of these sets are denoted as $|\mathcal{V}|=N$, $|\mathcal{E}|=E$ and $|\mathcal{T}|=T$.
Next, we introduce the concept of geometric simplicial complex  (GSC), of which   a triangular surface mesh is a typical example.\\ 
\textbf{Geometric Simplicial Complexes.}  Let us define a set of points $v_i$ in  $\mathbb{R}^d$  as affinely independent if it is not contained in a hyperplane.
 An affinely independent set in $\mathbb{R}^d$ contains at most $d+1$  points.
 A $k$-simplex $\sigma^k$
is the convex hull of $k + 1$ affinely independent points $\sigma^k=(v_0,\ldots,v_k)$.
  Thus, a  point   is a $0$-simplex $\sigma^0$, a line segment is a $1$-simplex $\sigma^1$, a triangle  is a $2$-simplex $\sigma^2$, and so on. A face of the $k$-simplex $\{v_{i_0},\ldots, v_{i_k}  \}$ is a
 $(k-1)$-simplex of the form $\{v_{i_0},\dots,v_{i_{j-1}},v_{i_{j-1}},\ldots,v_{i_k}\}$, for some $0 \leq j \leq k$.
  A  geometric simplicial complex $\mathcal{X}$ in $\mathbb{R}^d$   is a collection of simplices in $\mathbb{R}^d$ such that
i) every face of a simplex of $\mathcal{X}$ is in $\mathcal{X}$ and ii)
the  intersection of any two simplices of $\mathcal{X}$ is either a face of each of them, or it is empty \cite{hirani2003discrete}.  
 %$$\mathcal{T}=\left\{\sigma_k=\{\textbf{v}_{k_1},\textbf{v}_{k_2},\textbf{v}_{k_3}\},\; k=0,\cdots K-1\right\}.$$  The  triangles in $\mathcal{T}$ can be  identified by triangulation algorithms  (e.g. Delaunay \cite{lee1980two}, \cite{chen2004optimal}), and naturally induce the definition of a set of edges $\mathcal{E}$ built by the triangles edges; in formulas:
 %\begin{equation}
%\begin{split}
 % \mathcal{E}=\bigl\{
 % \epsilon_m=\%{\textbf{v}_{m_1},\textbf{v}_{m_2}\}, \;m=0,\cdots M-1
 % \bigr.
 % \\
 % \bigl.
 %\textbf{v}_{m_1},\textbf{v}_{m_2} \;\text{s.t.}\;\exists\; \overline{k}: \textbf{v}_{m_1},\textbf{v}_{m_2}\in \sigma_{\overline{k}}\; 
 %\bigr\}.   
 %\end{split}
%\end{equation}
We denote by $\sigma_i^{k}$ the simplex $i$ of order $k$.  
A $(k-1)$-face  $\sigma_j^{k-1}$
 of a $k$-simplex
$\sigma^k_i$
 is called a boundary element of $\sigma^k_i$. We use the notation $\sigma_j^{k-1} \prec_b \sigma^k_i$ to indicate that $\sigma_j^{k-1}$ is a simplex bounding $\sigma^k_i$. A $(k-1)$-simplex $\sigma_i^{k-1}$ is incident to a $k$-simplex $\sigma^{k}_j$ if  $\sigma^{k-1}_i \prec_b \sigma^{k}_j$.
Two simplices of order $k$ are lower incident if they share a common simplex of order $(k-1)$,  or upper incident  if they  are faces of a common simplex of order $k+1$. 
The structure of the complex ${\mathcal{X}}$ is described by the incidences matrices $\mB_k$ with $k=1,\ldots,K$, describing  which $k$-simplices are incident to which $(k-1)$-cells, and accounting for simplex orientation. These boundary matrices are defined as \cite{barbarossa2020topological}:
  \begin{equation}      
   \label{inc_coeff}
  B_k(i,j)=\left\{\begin{array}{rll}
    1,& \text{if} \; \sigma^{k-1}_i \prec_b \sigma^{k}_j \;  \text{and} \; \sigma^{k-1}_i \sim \sigma^{k}_j\\
  -1,& \text{if} \; \sigma^{k-1}_i\prec_b \sigma^{k}_j \;  \text{and} \; \sigma^{k-1}_i \nsim \sigma^{k}_j\\
  0, & \text{if} \; \sigma^{k-1}_i \not\prec_b \sigma^{k}_j \\
  \end{array}\right. 
\end{equation}
where we use the notation $\sigma^{k-1}_i \sim \sigma^{k}_j$ ($\sigma^{k-1}_i \nsim \sigma^{k}_j$) to indicate that the orientation of $\sigma^{k-1}_i$ is coherent with that of  $\sigma^{k}_j$. \\
Given a simplicial complex we can define the signals $\bs^0: \mathcal{V}\rightarrow \mathbb{R}^N$, $\bs^1: \mathcal{E}\rightarrow \mathbb{R}^E$ and $\bs^2: \mathcal{T}\rightarrow \mathbb{R}^T$ on the simplicial complex as the signals observed over the nodes, edges and triangles, respectively.
A geometric simplicial complex is a triangular mesh, as it encodes geometry through vertices, edges, and triangular faces. An example of a GSC is illustrated in Fig. \ref{fig:geom} where we show a color map on a spheroidal PC, with normal vectors associated to each triangle and colors associated to each simplex.
%Therefore, we associate the point cloud $\mathcal{P}$ with
%a simplicial complex $\mathcal{X}=(\mathcal{V},\mathcal{E},\mathcal{T})$.
%The simplex complex $\mathcal{X}$ is associated to the first order combinatorial Laplacian matrices %
%\cite{horak2013spectra} 
%given by
%\begin{equation}
%\begin{split}
%& \mL_0=\mB_1 \mB_1^T,\\
%&\mL_k=
%\underbrace{\mB_k^T\mB_k}_{\mL_{k,d}}
%+
%\underbrace{\mB_{k+1}\mB_{k+1}^T}_{\mL_{k,u}}\\
%\end{split}
%\label{eq:tsp}
%\end{equation}
%where $\mL_0$ is the (binary) graph Laplacian matrix and $\mL_{k,d}$ and $\mL_{k,u}$ denote  the lower and upper $k$-order Laplacians, expressing the lower and upper adjacencies of the $k$-order simplices, respectively. 
%In graph signal processing, the definition of the Laplacian $\mL_0$ can be modified, e.g. resorting to the Laplace-Beltrami algorithm to compute the Laplacian weights, so as to account for the graph geometry. Similarly, the  Laplacians describing the simplicial complex can be generalized to account of specific geometry properties, as discussed in the next sections. 

  \section{Topological Signals over Point Clouds}

%\begin{figure}
%    \centering    \includegraphics[width=0.7\linewidth]{figure_pc_grey.pdf}
%    \caption{Geometric  SC and observed vector field.}
%    \label{fig:geom}
%\end{figure}

Given a PC, the discrete vector field $\{\mathbf{v}(\mathbf{p}_i)\}_{i=0}^{N-1}$  is a map from  the point position $\mathbf{p}_i$ of $\mathcal{X}$ to $\mathbb{R}^3$. %, Associating each vertex to the vector of the corresponding color coordinates in a color space, e.g. in the RGB color space. 
The   vectors $\mathbf{v}(\mathbf{p}_i) \in \mathbb{R}^3,\; i=0,\cdots N-1$ can be processed applying topological signal processing \cite{barbarossa2020topological} and discrete exterior calculus  \cite{desbrun2005discrete}. 
We first project each node vector $\mathbf{v}(\mathbf{p}_i)$ on all the edges having node $i$ as an endpoint. More specifically, denoting by $m_1$, $m_2$ the endpoints of edge $m$, we build the scalar signal on edge $m$,  for $m=1,\ldots, E$,  as
\begin{equation} \label{eq:eq_sm}
s^{1}_m=
 \frac{1}{2} \left[
(\mathbf{p}_{m_2} - \mathbf{p}_{m_1})^\top 
\mathbf{v}(\mathbf{p}_{m_1}) +
(\mathbf{p}_{m_2} - \mathbf{p}_{m_1})^\top 
\mathbf{v}(\mathbf{p}_{m_2}) \right].
\end{equation}

%\begin{equation} \label{eq:eq_sm}
%s^{1}_m=
%  \sum_{i \in \mathcal{N}_{m_1,m_2}}\frac{1}{2}
% (\mathbf{p}_{m_2} - \mathbf{p}_{m_1})^\top 
% \mathbf{v}(\mathbf{p}_{i})  
%\end{equation}

%\begin{align}
%\label{eq:eq_sm}
%s^{1}_m =
%\sum_{m_1,m_2 \in \mathcal{N}_{m_1,m_2}} \frac{1}{2}
%\Big[
%&(\mathbf{p}_{m_2} - \mathbf{p}_{m_1})^\top 
%\mathbf{v}(\mathbf{p}_{m_1}) \nonumber \\
%&+ (\mathbf{p}_{m_2} - \mathbf{p}_{m_1})^\top 
%\mathbf{v}(\mathbf{p}_{m_2})
%\Big]
%\end{align}

%where $\mathcal{N}_{m_1,m_2}$ is the set of nodes neighboring  the vertices $m_1$ and $m_2$.
Thus, we obtain a scalar field defined over the edges of the simplicial complex. Next, we use Whitney interpolation \cite{dodziuk1976finite} to compute the vector associated to the barycenter of each triangle, as follows. Let us define by  $\mathbf{v}(\mathbf{p};\sigma^2)=\mathbf{T}_{\sigma}\mathbf{v}(\mathbf{p})$ the  vector at position $\mathbf{p}$ tangential to the triangle $\sigma^2$, where $\mathbf{T}_{\sigma}=\mI-\mathbf{n}_{\sigma}\mathbf{n}^T_{\sigma}$ with $\mathbf{n}_{\sigma}$  the normal to the triangular face. According to Whitney interpolation,  we reconstruct from the edge signal in (\ref{eq:eq_sm}) the field $\mathbf{v}(\mathbf{p};\sigma^2)$ as
\begin{equation} \label{eq:Rec_v}
\begin{split}
\mathbf{v}(\mathbf{p}; \sigma^2)& = 
\sum\limits_{m=1}^{E} s^{1}_{m}
\cdot
\big[\varphi^{(2)}_{m_1}(\mathbf{p})
\nabla  \varphi^{(2)}_{m_2}(\mathbf{p})- \\
&\varphi^{(2)}_{m_2}(\mathbf{p})\nabla \varphi^{(2)}_{m_1}(\mathbf{p}) \big],\;\mathbf{p}\subset\sigma^2%\;\sigma^2\in\mathcal{T}_{m_1}, \mathcal{T}_{m_2}
\end{split}
\end{equation}
where $\varphi^{(2)}_i(\mathbf{p})$ is an affine piecewise function, belonging to the class of Whitney interpolation functions and $\sigma^2 \in \mathcal{T}_i$, with  $\mathcal{T}_i$  the subset of triangles to which the i-vertex belongs to.

%and 
%$\sigma^2$ belongs to $\mathcal{T}_i$ the set of triangles incident to nodes  or 
The  Whitney interpolation functions of order $k$ are supported  over $k$-simplices \cite{desbrun2005discrete}. The function $\varphi_i^{(2)}(\mathbf{p})$ is supported on the $2$-simplices  having one vertex in $\mathbf{p}_i$, i.e. for each $\mathbf{p}$ in  $\mathcal{T}_i$. On each of the $2$-simplices,  $\varphi^{(2)}_i(\mathbf{p})$ takes value $1$ at $\mathbf{p}_i$ and decreases linearly to zero at  the other  vertices. 
% per nostra referenza https://mdav.ece.gatech.edu/ece-6250-fall2019/notes/12-notes-6250-f19.pdf
%1)Linear Algebra and its Applications Projections under seminorms and generalized Moore Penrose inverses  Mitra ∗Rao ∗
%.2) R. Rao and S. K. Mitra, Generalized Inverse of Matrices and its Applications, Wiley, New York (1971b).

The final step consists in evaluating the vector field over the vertices of the original point cloud. To do that, we first derive all the vectors on the barycenters  $\mathbf{p}_\sigma$ of each triangle, 
setting $\mathbf{p}=\mathbf{p}_\sigma$ in  (\ref{eq:Rec_v}). In this manner, the tangential vector field computed over all the barycenters is $\mathbf{v}_i(\sigma^2)= \mathbf{v}(\mathbf{p}; \sigma^2)|_{\mathbf{p}=\mathbf{p}_{\sigma^2}}$.
Given the collection of   tangential components $\mathbf{
v}_i(\sigma^2), \;\sigma^2\in\mathcal{T}_i$,   the  discrete vector field $\mathbf{v}_i$ at the  $i$-th graph vertex is reconstructed in closed-form as  \cite{rao1972generalized} \cite{mitra1974projections}:  
\begin{equation}
\mathbf{v}_i = \left(\sum_{\sigma^2 \in \mathcal{T}_i} \mathbf{T}_\sigma^\top \mathbf{T}_\sigma\right)^{-1} \sum_{\sigma^2 \in \mathcal{T}_i} \mathbf{T}_\sigma^\top \mathbf{v}_i(\sigma^2)
\end{equation}
provided that the signal  $\mathbf{v}_i$ is not perpendicular to all the incident faces, i.e.  $\exists \;\sigma^2\in\mathcal{T}_i|\mathbf{v}_i\not\perp\sigma^2$.
%This is detailed in the following subsection, where we show how the reconstructed signal on graph $\mathbf{v}_i,\; i=0,\cdots N-1$ is consistent with the  edge signal $s^{1}_m,\; m=1,\cdots E$, enabling the round-trip mapping.

Finally, the adoption of the Whitney interpolation function to reconstruct the vector field over the 2-simplices allows to endow the  above defined notion  of incidence between simplices with a geometric metric. This leads to the  definition of  the first-order discrete  Laplacian $\textbf{L}_1$, weighted with the metric matrices $\mathbf{M}_p$ as \cite{fisher2007design}
\begin{equation} \label{eq:L_1}
{\textbf{{L}}}_1 = \mathbf{M}_1 \mB_1^T \mathbf{M}_0^{-1} \mB_1 \mathbf{M}_1 +\mB_2 \mathbf{M}_2 \mB_2^T
\end{equation}
where the weighting matrices $\textbf{M}_0$, $\textbf{M}_1$ and $\textbf{M}_2$, depend on 
the forms used  for vector field interpolation \cite{bell2012pydec}. Using the 
Whitney interpolation functions on $k$-simplices, the metric matrices $\textbf{M}_k$ define Whitney inner products between signals, with  entries given by
%Using the inner product of  Whitney interpolation functions over $k$-simplices, we get:
\begin{equation}
|\textbf{M}_k|_{ij} = \langle \varphi_i^{(k)}, \varphi_j^{(k)} \rangle, \quad k=0,1,2.
\end{equation}
The weighted first-order Laplacian in (\ref{eq:L_1}) is composed of two terms: the first term $\mathbf{L}_{1,d}=\mathbf{M}_1 \mB_1^T \mathbf{M}_0^{-1} \mB_1 \mathbf{M}_1$ describes the lower adjacencies between edges, while the second term $\mathbf{L}_{1,u}=\mB_2 \mathbf{M}_2 \mB_2^T$ described the upper adjacencies of edges as boundaries of triangles.
Then, $\mathbf{L}_1$ provides an algebraic representation of the triangular mesh enabling the use of the topological signal processing framework \cite{barbarossa2020topological} for processing signals observed over the edges of a GSC. Specifically, denoting with  $\mU=\{\textbf{u}_m\}_{m=1}^{E}$  the eigenvectors of the first order Laplacian matrix $\mL_1$,   the Simplicial Fourier Transform (SFT) $\hat{\mathbf{s}}^{1}$ of the edge signal ${\mathbf{s}}^{1}$ is defined as $\hat{\mathbf{s}}^{1}=\mU^T \mathbf{s}^{1}$. This transform provides  the spectral representation of the edge signals as ${\mathbf{s}}^{1}=\mU^T \hat{\mathbf{s}}^{1}$.

\begin{figure}[htb]
    \centering
    \includegraphics[width=0.8\linewidth]{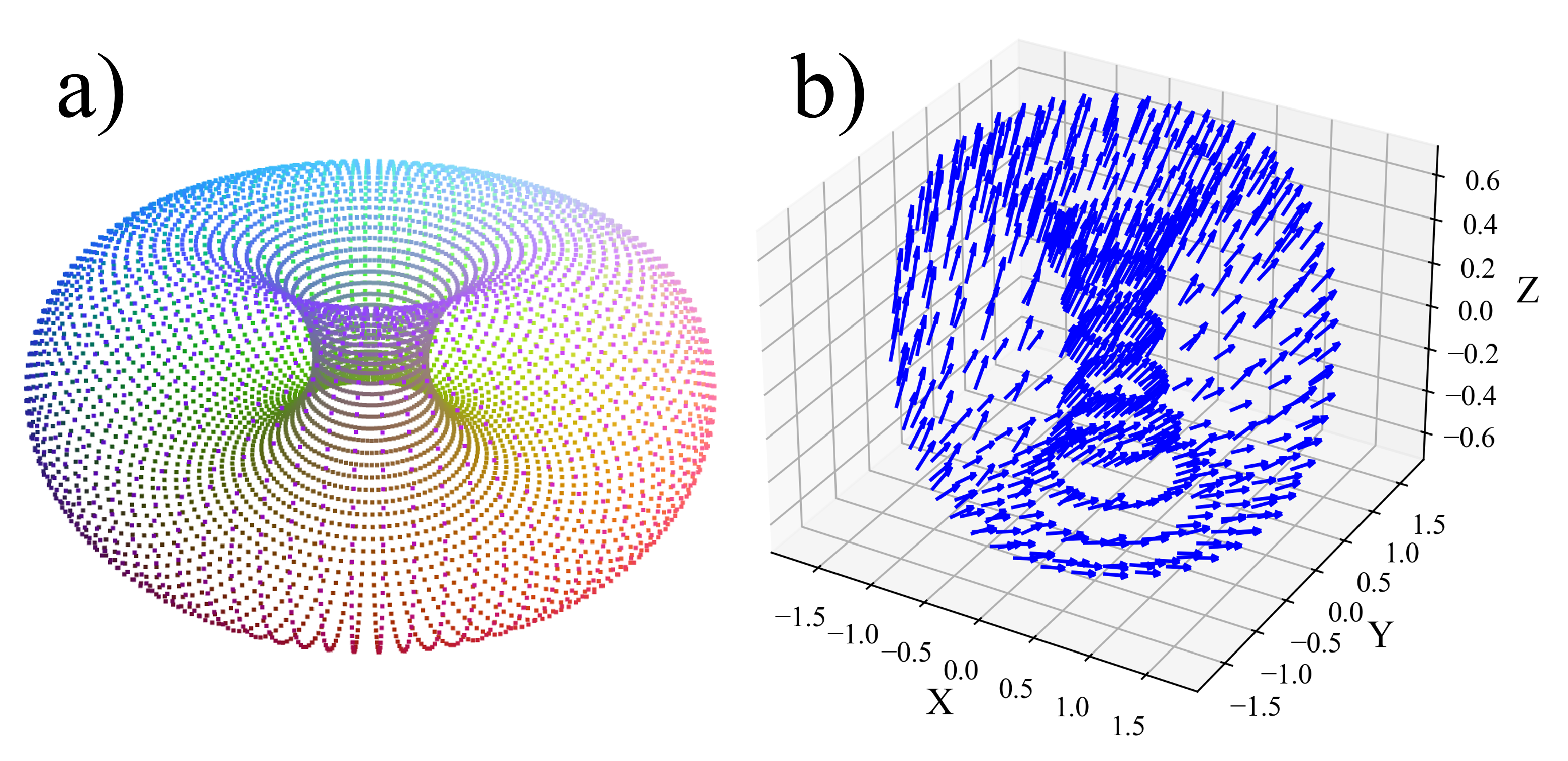}
    \caption{(a) Toroidal PC  and (b) associated discrete color vector field.}
    \label{fig:pc}
\end{figure}
\section{Point Cloud Color Recovery}
Let us now consider the vector field of the colors observed over a PC, i.e. a map from  the point position $\mathbf{p}_i$ of $\mathcal{X}$ to $[0,1]^3$. 
In the previous section, we showed how using (\ref{eq:Rec_v}) we can recover the point cloud color, represented as a discrete vector field, from scalar signals on the edges $s^{1}_m,\; m=1,\ldots, E$. 
However, in many practical applications, color attributes may be missing for some points due to sensor errors, or intentionally subsampled during compression to reduce the bit rate. 
In such cases, the scalar signal $s^{1}_m$ is unavailable on some edges. 
We address the problem of  recovering   $s^{1}_m,\; \forall m$, from its values on a subset ${\cal S}$ of observed edges, assuming $|{\cal S}|< E$. 
To identify the optimal sampling set we use the greedy MaxDet strategy in \cite{tsitsvero2016signals}.
Let us express the sampled edge signal in vector form as:
\begin{equation} \my^{1}=\mD_{\cal S}\, \mathbf{s}^{1} \end{equation}
where  $\mathbf{s}^{1}=\big[s_1^{1},\cdots s_{E}^{1}\big]^T$  and  $\mD_{\cal S}$ is an edge-selection diagonal matrix whose $m$-th diagonal entry is  $1$ if $m \in {\cal S}$, and $0$ otherwise.
The conditions for recovering the edge  signal ${\mathbf{s}}^{1}$ from its edge samples in ${\cal S}$ are given in \cite{barbarossa2020topological}, \cite{tsitsvero2016signals} in terms of its  SFT \cite{barbarossa2020topological}. 

Let us assume that the edge signal  ${\mathbf{s}}^{1}$   is $\mathcal{K}$-bandlimited, i.e. it can be represented over  $K$ eigenvectors $\mU_{\mathcal{K}}=\{\mathbf{u}_i\}_{i\in \mathcal{K}}$ of frequency indexes $i\in\mathcal{K}$, i.e.  $\mathbf{s}^{1}=\mU_{\mathcal{K}} \hat{\mathbf{s}}^{1}_{\mathcal{K}}$. With these positions, the edge signal  ${\mathbf{s}}^{1}$   is recovered from $N_{sc}=|\mathcal{S}|$ samples with $N_{sc}>K$   \cite{tsitsvero2016signals} as follows:
\begin{equation}    
\label{eq:interpol}
\mathbf{s}^{1}=  \big[\mI-\big(\mI-{\mD}_{\cal S}\big) \mU_{\mathcal{K}} \mU_{\mathcal{K}}^T\big]^{-1} \my^{1}.
\end{equation}
 
\begin{figure}
    \centering
    \includegraphics[width=0.7\linewidth]{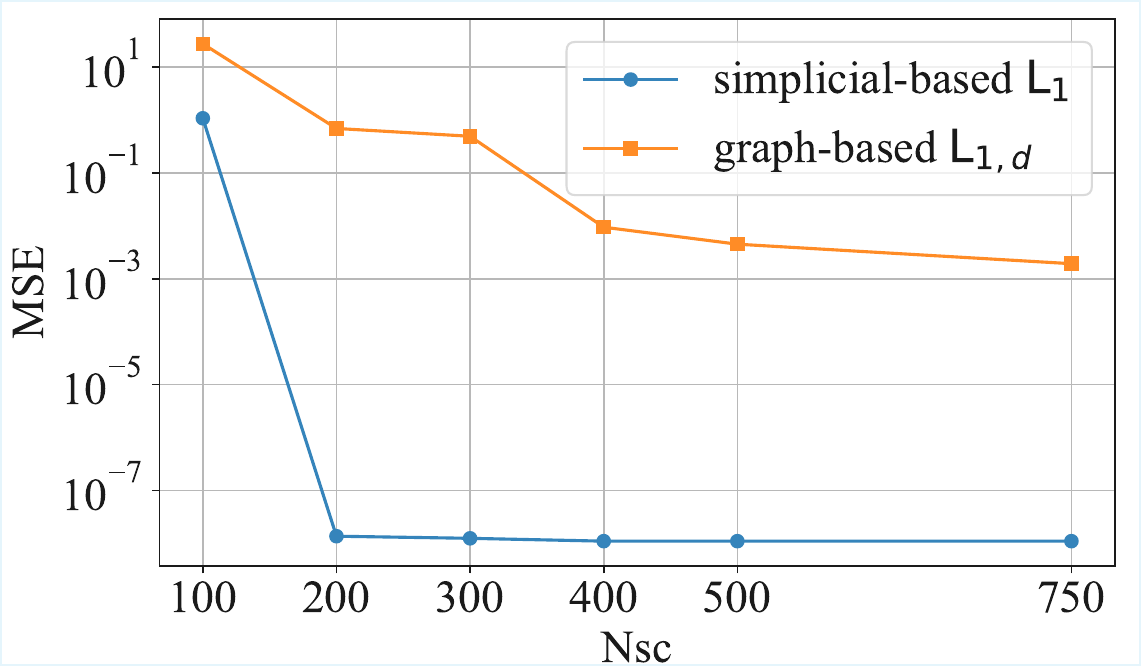}
    \caption{MSE versus observed samples.}
    \label{fig:mse}
\end{figure}

We  assess the reconstruction performance of the proposed  approach on a toroidal PC,  with $N = 600$ points of coordinates bounded in $[0, x_{\max}]\times[0, y_{\max}]\times[0, z_{\max}]$. The associated triangular mesh was built via a standard parametric tessellation, leading to $E = 1800$ edges  and $T = 1200$ triangles.  The point cloud color $[R_i,G_i, B_i]$ at node $i$ is assigned  as $R_i=\dfrac{x_i}{x_{\max}}$, $G_i=\dfrac{y_i}{y_{\max}}$, $B_i=\dfrac{z_i}{z_{\max}}$.  This setup ensured a smooth color distribution over the point cloud that we used as the ground truth signal.  Specifically, the point cloud color attribute is represented by the discrete vector field $\mathbf{v}_i=[R_i,G_i,B_i]^T,\;i=0,\cdots N-1$. 
Figure \ref{fig:pc} represents the colored point cloud and the quiver plot of the discrete vector field $\mathbf{v}_i,\;i=0,\cdots N-1$. 
Then,  using (\ref{eq:interpol}) we recover   the edge signal   $\mathbf{s}^{1}$ from its values collected over a subset ${\cal S}$ of edges with $N_{sc}<E$.   
We compared two reconstruction strategies:  
(i) the graph-based reconstruction relying only on the edge Laplacian $\mathbf{L}_{1,d}$ (without considering triangles), and  (ii) the proposed simplicial-complex reconstruction using the full Hodge Laplacian $\mathbf{L}_1$.  
The sampling and reconstruction process was carried out by varying the number of sampled nodes $N_{sc}$ from $100$ to $750$ and recovering the edge signal $\hat{\mathbf{s}}^{1}$. In Fig. \ref{fig:mse}, we report the mean square error $\text{MSE}=\parallel \hat{\mathbf{s}}^{1}-{\mathbf{s}}^{1} \parallel$  between the reconstructed and original signals. 
The results show how the use of simplicial complexes significantly improves the accuracy of the reconstruction with respect to graph-based representation. This confirms that incorporating higher-order topological information provides a richer basis for signal representation, enabling a more accurate recovery of the underlying edge signals from sparse samples.

% \begin{figure}
%     \centering
%     \includegraphics[width=0.95\linewidth]{pc_orig.png}
%     \caption{Toroidal PC and the associated discrete color vector field.}
%     \label{fig:pc}
% \end{figure}

\section{Point Cloud Geometry Refinement}
\begin{figure}
    \centering
    \includegraphics[width=0.7\linewidth]{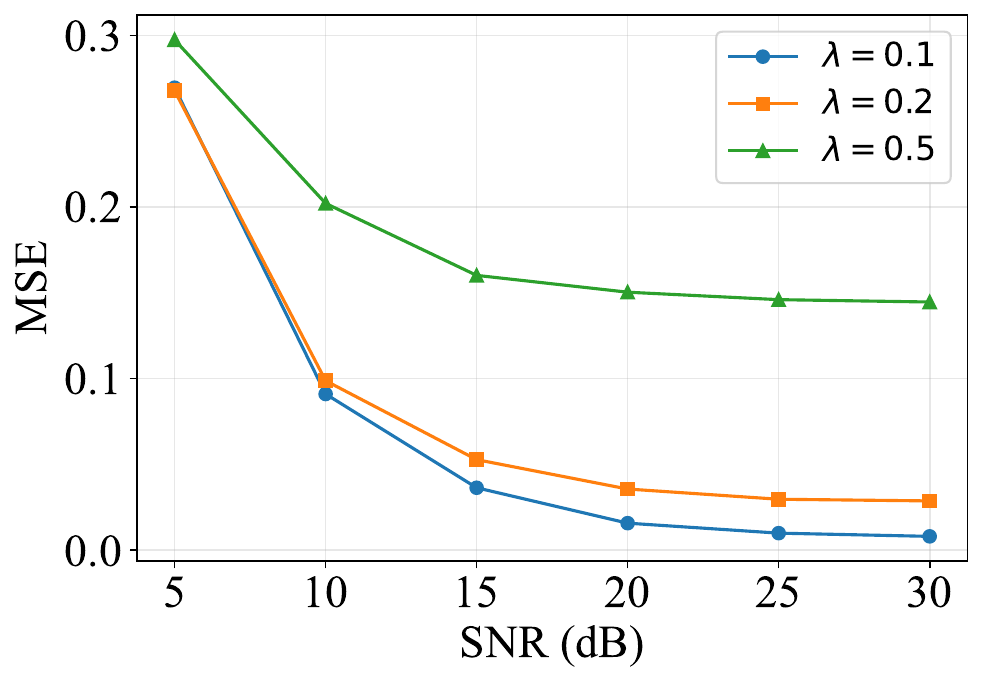}
    \caption{MSE of the denoised normals with respect to the noiseless case, as a function of the SNR.}
    \label{fig:msesnr}
\end{figure}

\begin{figure}
    \centering
    \includegraphics[width=0.82\linewidth]{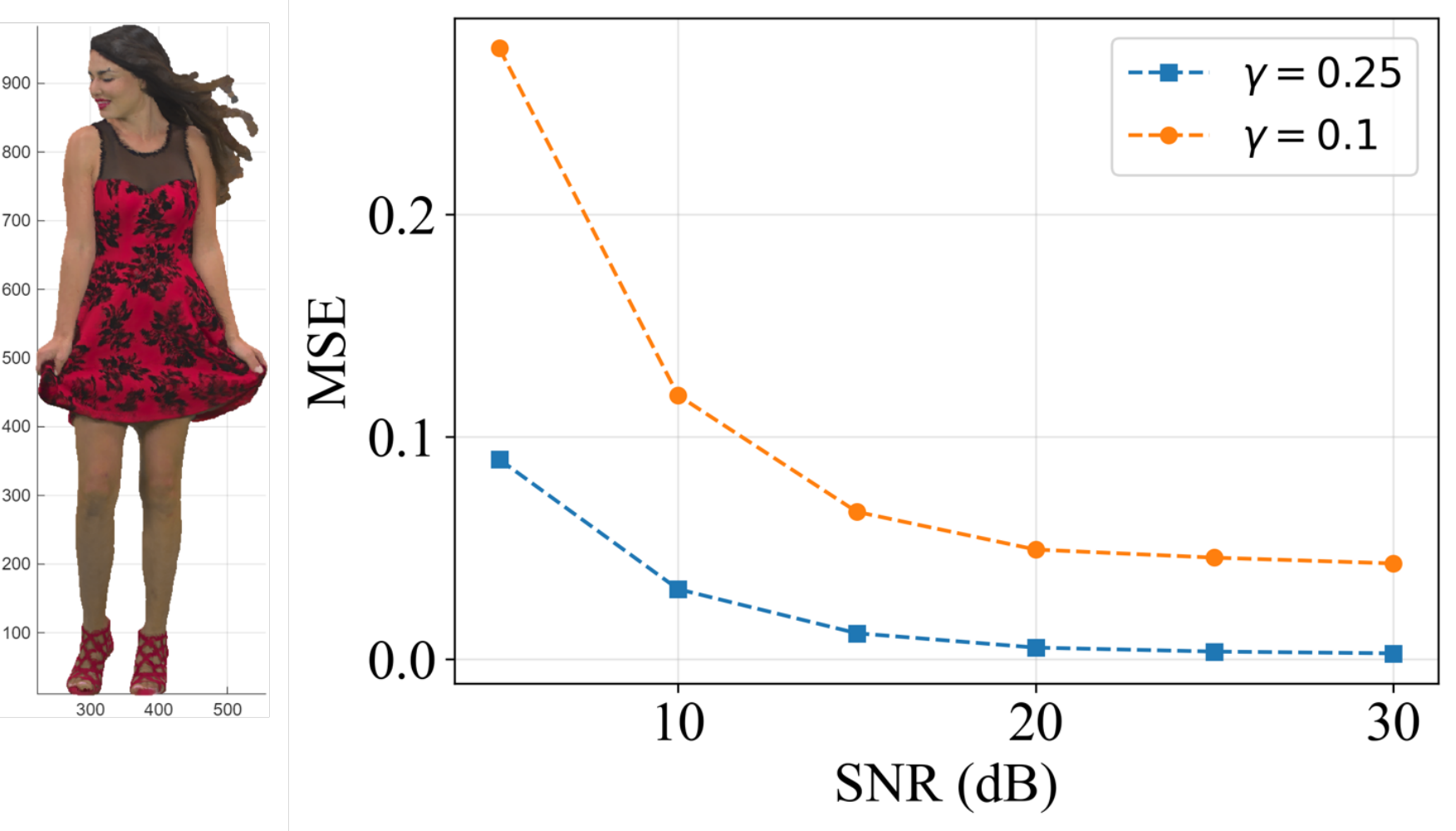}
    \caption{MSE of the denoised normals as a function of the SNR for Red and Black PC varying $\gamma$ with a fixed $\lambda=0.1$. }
    \label{fig:realPC}
\end{figure}

When capturing the PC from sensors like LiDAR or lasers, the coordinates of the points may be affected by errors. This geometric perturbation translates into noise on the normals of the planes defined by the mesh triangles. To deal with this kind of perturbation, we treat the vector associated with the normals as a signal defined over the triangles of the simplicial complex. The problem can then be cast as a geometry noise filtering, involving the weighted the second-order Hodge Laplacian matrix \cite{hirani2010cohomologous}
\begin{align}
{\mathbf{{L}}}_2 & = \mathbf{M}_2 \mB_2^T \mathbf{M}_1^{-1} \mB_2 \mathbf{M}_2.
\label{eq:Ldue}
\end{align}
Specifically, we formulate the geometry refinement as the following optimization problem:
\begin{equation}
\min_{\mathbf{s}^{2} \in \mathbb{R}^T} 
\; \| \mathbf{s}^{2} - \mathbf{x}^{2} \|_2 
+ \lambda \, {\mathbf{s}^{2}}^\top \mathbf{{L}}_2 \mathbf{s}^{2} 
+ \gamma \, \| \mathbf{s}^{2} \|_1
\end{equation}
where $\mathbf{x}^{2}$ are the noisy signals given by the normals to the triangular faces, while
$\mathbf{s}^2$ is the filtered estimate signal. 
Our goal is to minimize a weighted sum of data fitting error,  signal smoothness, and sparsity. Specifically, 
the first term of the objective function is the data fitting error, the second term represents the smoothness of the signal over the second-order Hodge Laplacian  $\mathbf{L}_2$, while the third term is introduced to control the signal sparsity. 
The coefficients $\gamma$ and $\lambda$ are tuned to control the trade-off between the objective function terms. 
We tested the proposed algorithm on the same PC described in the previous section to denoise triangular signals on simplicial complexes, i.e. the normals to the triangles.  
Specifically, we add Gaussian noise to the normals, which corresponds to perturbing the geometry of the PC. %In Fig.~\ref{fig:norm} we illustrate the recovered normals (black quivers) and the noisy normal (red quivers). 
In Fig.~\ref{fig:msesnr}, we show the MSE of the denoised normals with respect to the noiseless case, as a function of the signal-to-noise ratio (SNR). 
The results are obtained by fixing 
$\gamma = 0.1$ and varying $\lambda$ in the set $\{0.1,\,0.2,\,0.5\}$, leading to three different curves. As expected, the curves show an improvement in performance with increasing SNR. Furthermore, a smaller $\lambda$ leads to a lower MSE, as  the data fitting term prevails in the optimization. We also test our method on a real PC (Red and Black) \cite{d20178i} and we show the ability of the proposed approach to refine geometry in presence of additive noise in Fig. \ref{fig:realPC}.

\section{Conclusions}
In this paper we employed the topological signal processing framework for processing a PC, modeled as a vector field defined on a set of points whose coordinates may be affected by errors. 
We showed that leveraging the higher order structure improves color recovery over graph-based methods, and that geometry refinement can be effectively performed. 
Of course, there is a lot to be done to extend our approach to cases where the filtering operator and the geometry refinement can be implemented jointly. A very interesting aspect to be investigated is the use of space-varying filtering operator that adapts the support of its kernel to the curvature of the point cloud.  

\vfill\pagebreak

%\section{REFERENCES}
%\label{sec:refs}

\bibliographystyle{IEEEbib}
\bibliography{biblio}

\end{document}